\documentclass[aps,prb,twocolumn,superscriptaddress]{revtex4}
\usepackage{amsmath,amssymb,color,graphicx,dcolumn,bm}

\usepackage{lineno}
\usepackage{graphicx}
\usepackage{color}
\usepackage{amssymb}
\usepackage{amsmath}
\usepackage{dsfont}
\usepackage{comment}
\usepackage{color,soul}
\usepackage{hyperref}
\hypersetup{colorlinks=true,
linkcolor=red,
anchorcolor=blue,
citecolor=blue,
filecolor=blue,
menucolor=blue,
urlcolor=blue
}

\newcommand{\bra}[1]{\ensuremath{\langle{#1}|}}
\newcommand{\ket}[1]{\ensuremath{|{#1}\rangle}}

\newcommand{\abs}[1]{\ensuremath{\left|{#1}\right|}}

\def\ba{\begin{eqnarray}}
\def\ea{\end{eqnarray}}
\def\beq{\begin{equation}}
\def\eeq{\end{equation}}
\def\beqstar{\begin{equation*}}
\def\eeqstar{\end{equation*}}

\begin{document}
\title{Observation of discrete time-crystalline order in a disordered dipolar many-body system}


\author{Soonwon Choi}
\thanks{These authors contributed equally to this work.}
\affiliation{Department of Physics, Harvard University, Cambridge, Massachusetts 02138, USA}
\author{Joonhee Choi}
\thanks{These authors contributed equally to this work.}
\affiliation{Department of Physics, Harvard University, Cambridge, Massachusetts 02138, USA}
\affiliation{Harvard John A. Paulson School of Engineering and Applied Sciences, Harvard University, Cambridge, Massachusetts 02138, USA}
\author{Renate Landig}
\thanks{These authors contributed equally to this work.}
\affiliation{Department of Physics, Harvard University, Cambridge, Massachusetts 02138, USA}
\author{Georg Kucsko}
\affiliation{Department of Physics, Harvard University, Cambridge, Massachusetts 02138, USA}
\author{Hengyun Zhou}
\affiliation{Department of Physics, Harvard University, Cambridge, Massachusetts 02138, USA}
\author{Junichi Isoya}
\affiliation{Research Centre for Knowledge Communities, University of Tsukuba, Tsukuba, Ibaraki 305-8550, Japan}
\author{Fedor Jelezko}
\affiliation{Institut f{\"u}r Quantenoptik, Universit{\"a}t Ulm, 89081 Ulm, Germany}
\author{Shinobu Onoda}
\affiliation{Takasaki Advanced Radiation Research Institute, National Institutes for Quantum and Radiological Science and Technology, 1233 Watanuki, Takasaki, Gunma 370-1292, Japan}
\author{Hitoshi Sumiya}
\affiliation{Sumitomo Electric Industries Ltd., Itami, Hyougo, 664-0016, Japan}
\author{Vedika Khemani}
\affiliation{Department of Physics, Harvard University, Cambridge, Massachusetts 02138, USA}
\author{Curt von Keyserlingk}
\affiliation{Princeton Center for Theoretical Science, Princeton University, Princeton, New Jersey 08544, USA}
\author{Norman Y. Yao}
\affiliation{Department of Physics, University of California Berkeley, Berkeley, California 94720, USA}
\author{Eugene Demler}
\affiliation{Department of Physics, Harvard University, Cambridge, Massachusetts 02138, USA}
\author{Mikhail D. Lukin}
\affiliation{Department of Physics, Harvard University, Cambridge, Massachusetts 02138, USA}

\maketitle

\textbf{
Understanding quantum dynamics away from equilibrium is an outstanding challenge in the modern physical sciences. 
It is well known that out-of-equilibrium systems can display a rich array of phenomena, ranging from self-organized synchronization to dynamical phase transitions~\cite{adler1946study,Cross1993pattern}. 
More recently, advances in the controlled manipulation of  isolated many-body systems have enabled detailed studies of non-equilibrium phases in strongly interacting quantum matter~\cite{schreiber2015observation, langen2015experimental,monroe2016MBL,kaufman2016quantum}. 
As a particularly striking  example, the interplay of periodic driving, disorder, and strong interactions has recently been predicted to result in exotic ``time-crystalline'' phases \cite{wilczek2012quantum,li2012}, which spontaneously break the discrete time-translation symmetry of the underlying drive~\cite{vedika2016phase,else2016floquet,von2016absolute,yao2016discrete}.
Here, we report the experimental observation of such discrete time-crystalline order in a driven, disordered ensemble of $\sim 10^6$ dipolar spin impurities in diamond at room-temperature~\cite{jelezko2004observation,Childress281,NVDD,Doherty20131}. We observe long-lived temporal correlations at integer multiples of the fundamental driving period,
experimentally identify the phase boundary and find that the temporal order is protected by strong interactions; this order is remarkably stable against perturbations, even in the presence of slow thermalization~\cite{anderson1958absence, kucsko2016critical}. 
Our work opens the door to exploring dynamical phases of matter and controlling interacting, disordered many-body systems~\cite{deutsch2010spin, anaMaria2008,cappellaro2009quantum}. 
}

Conventional wisdom holds that the periodic driving of isolated, interacting systems inevitably leads to heating and the loss of quantum coherence. 
In certain cases, however, fine-tuned driving can actually decouple quantum degrees of freedom from both their local environment \cite{NVDD} and from each other \cite{waugh1968approach}.
Recently, it has been shown that strong disorder, leading to the so-called many-body localization (MBL) \cite{basko2006metal,HuseMBLreview}, allows the systems to retain  memory of their initial state for long times, enabling the observation of novel, out-of-equilibrium quantum phases \cite{schreiber2015observation, monroe2016MBL,abanin2015exponentially}. %
One example is the discrete time crystal (DTC)~\cite{vedika2016phase,else2016floquet,von2016absolute,yao2016discrete}, a phase which is nominally forbidden in equilibrium \cite{bruno2013impossibility, Watanabe2015}. 
The essence of the DTC phase is an emergent, collective, subharmonic temporal response~\cite{von2016absolute}.
While this phenomenon resembles the coherent revivals associated with dynamical decoupling~\cite{Childress281}, its nature is fundamentally different as it is induced and protected by  interactions rather than fine-tuned control fields.
It is especially intriguing to investigate the possibility of DTC order in systems that are not obviously localized~\cite{else2016pre}.
This  is the case for dipolar spins in three dimensions, where the interplay between interactions and disorder
can lead to critical sub-diffusive dynamics \cite{kucsko2016critical,choi2016depolarization}.

\begin{figure}[ht]
\includegraphics[width=76mm]{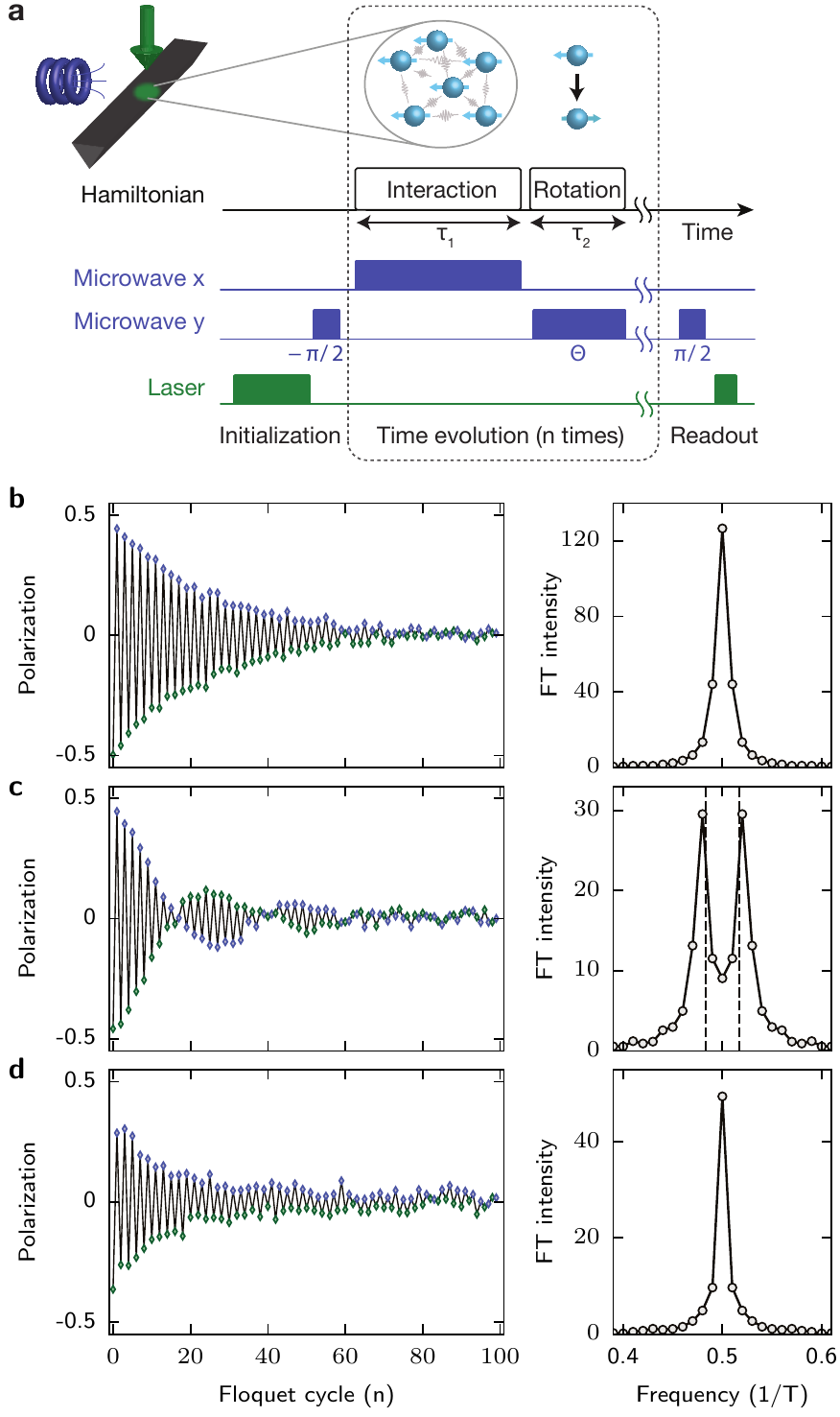}
\caption{\textbf{Experimental setup and sequence for observing time-crystalline order.}
\textbf{a}, NV centers in a nanobeam fabricated from black diamond are illuminated by a focused green laser beam and irradiated by a microwave source.
Within one Floquet cycle, the spins evolve under a dipolar interaction for duration $\tau_1$, followed by a global spin rotation acting for duration $\tau_2$. Experimental sequence: spins are prepared in the $(\ket{m_s=0}+\ket{m_s=-1})/\sqrt{2}$ state using a microwave $(-\pi/2)$-pulse along the $\hat{y}$ axis.
Subsequently, the spins  evolve for $\tau_1$ under a strong microwave field aligned along the $\hat{x}$ axis, immediately followed by a strong microwave $\theta$-pulse along the $\hat{y}$ axis.
After $n$ repetitions of the Floquet cycle, the spin polarization is read out by applying another microwave $(\pi/2)$-pulse along the $\hat{y}$ axis.    
\textbf{b-d}, Representative time traces of the spin polarization $P(nT)$ and respective Fourier spectra for different values of interaction time $\tau_1$ and $\theta$: (\textbf{b}) $\tau_1=92$~ns, $\theta=\pi$, (\textbf{c}) $\tau_1=92$~ns, $\theta=1.034 \pi$, and (\textbf{d}) $\tau_1=989$~ns, $\theta=1.034\pi$.
Data are averaged over more than $2\cdot10^4$ measurements.
Dashed lines in \textbf{c} indicate $\nu = \pm\theta/2\pi$.
}
\label{fig:Figure1}
\end{figure}

We experimentally investigate the formation of discrete time-crystalline order in an ensemble of nitrogen vacancy (NV) spin impurities in diamond.
Each NV center has an electronic  $S=1$ spin, from which we isolate an effective two level system by applying an external magnetic field.
These isolated spin states can be optically initialized/detected and  manipulated via microwave radiation~\cite{jelezko2004observation,Childress281,Doherty20131} (see Fig.~1a and Methods).
Our sample has a high concentration (45~ppm) of NV centers, giving rise to strong long-range magnetic dipolar interactions~\cite{kucsko2016critical}.
The spins are also subject to multiple sources of disorder owing to lattice strain, paramagnetic impurities and the random positioning of NV centers. 
 A strong, resonant microwave field is used  to control spin orientations, resulting in an effective Hamiltonian (in the rotating frame),\cite{kucsko2016critical}
\begin{align}
    H(t) =& \sum_i\Omega_x(t)  S_i^x + \Omega_y (t)  S_i^y +\Delta_i S_i^z \nonumber\\
    &+ \sum_{ij} (J_{ij}/r_{ij}^3) ( S_i^x S_j^x + S_i^y S_j^y - S_i^z S_j^z).
   \label{eqn:ham}
\end{align}
Here, $S_i^\mu$ ($\mu \in \{x, y, z\}$) are Pauli spin-$1/2$ operators acting on the effective two-level system spanned by the spin states $\ket{m_s=0}$ and $\ket{m_s=-1}$, $\Omega_{x(y)}$ is the Rabi frequency of the microwave driving, $\Delta_i$ is a disordered on-site  field with  approximate standard deviation $W=2\pi\times4.0~$MHz, $r_{ij}$ is the distance between spins $i$ and $j$ (average nearest-neighbor separation $r_0 \sim 8$~nm), and $J_{ij}$ are the orientation dependent coefficients of the dipolar interaction. We note that the average interaction, $J_{ij}/r_0^3 \sim 2\pi\times 105~$kHz~\cite{kucsko2016critical}, is significantly faster than typical spin coherence times~\cite{Doherty20131}.

In order to probe the existence of time-crystalline order, we monitor the spin dynamics of an initial state polarized along the $+\hat{x}$ direction.
We begin by applying continuous microwave driving (spin locking) along $\hat{x}$ with Rabi frequency $\Omega_x =  2\pi \times54.6$~MHz  for a duration $\tau_1$ (Fig.~1a). 
Next, we rotate the spin ensemble by an angle $\theta$ around the $\hat{y}$~axis using a strong microwave pulse with $\Omega_y = 2\pi \times 41.7$~MHz  for duration $\tau_2 = \theta / \Omega_y \ll \tau_1$.
This two-step sequence defines a Floquet unitary with a total  period $T = \tau_1 + \tau_2$ and is repeated $n$ times,
before the polarization $P(nT)$ along the $\hat{x}$~axis is measured. 
The resulting polarization dynamics are analyzed in both the time and frequency domain.
Repeating these measurements with various values of $\tau_1$ and $\theta$ allows  us to independently explore the effect of interactions and global rotations. 
We note that $\tau_1$  is chosen as an integer multiple of $2\pi/\Omega_x$ in order to avoid a self-correcting  dynamical decoupling~\cite{NVDD}.

Figure 1b-d depicts representative time traces and the corresponding Fourier spectra, $S(\nu) \equiv \sum_n P(nT) e^{i 2\pi n \nu}$, for various values of $\tau_1$ and $\theta$.
For relatively short interaction time $\tau_1 = 92$~ns and nearly perfect $\pi$-pulses ($\theta \approx \pi$), we observe that the spin polarization $P(nT)$ alternates between positive and negative values, resulting in a sub-harmonic peak at $\nu = 1/2$ (Fig.~1b). In our experiment, the microwave pulses have an intrinsic uncertainty $\sim1\%$ stemming from a combination of spatial inhomogeneity in the microwave fields, on-site potential disorder, and the effect of dipolar interactions (see Methods). These eventually cause the oscillations to decay after $\sim50$ periods.
While such temporal oscillations nominally break discrete time-translation symmetry, their physical origin is trivial.
To see this, we note that for sufficiently strong microwave driving, $\Omega_x \gg W, J_{ij}/r_0^3$,  the dynamics during $\tau_1$  are governed by an effective polarization-conserving Hamiltonian~\cite{kucsko2016critical}, $H_\textrm{eff} \approx  \sum_i \Omega_x S_i^x + \sum_{ij} (J_{ij}/r_{ij}^3)  S_i^x S_j^x$. 
During $\tau_2$, the evolution can be approximated as $R_y^\theta \approx e^{-i \theta  \sum_i S_i^y}$. When $\theta = \pi$, this pulse simply flips the sign of the $\hat{x}$ polarization during each Floquet cycle, resulting in the $\nu=1/2$ peak. However, this $2T$-periodic response originates from the fine tuning of $\theta$ and should not be robust against perturbations. 
Indeed, a systematic change in the average rotation angle to $\theta = 1.034  \pi$ causes the $2T$-periodicity to completely disappear, resulting in a modulated, decaying signal with two incommensurate Fourier peaks at $\nu = \pm\theta/2\pi$ (Fig.~1c). Remarkably, we find that a rigid $2T$-periodic response is restored when interactions are enhanced by increasing $\tau_1$  to 989~ns, suggesting that the $\nu=1/2$ peak is stabilized by  interactions.
In this case, we observe a sharp subharmonic peak in the spectrum at $\nu = 1/2$ and the oscillations in $P(nT)$  continue beyond $n\sim 100$ (Fig.~1d).  We associate this with DTC order \cite{vedika2016phase,else2016floquet,von2016absolute,yao2016discrete}.


\begin{figure}[ht]
\includegraphics[width=76mm]{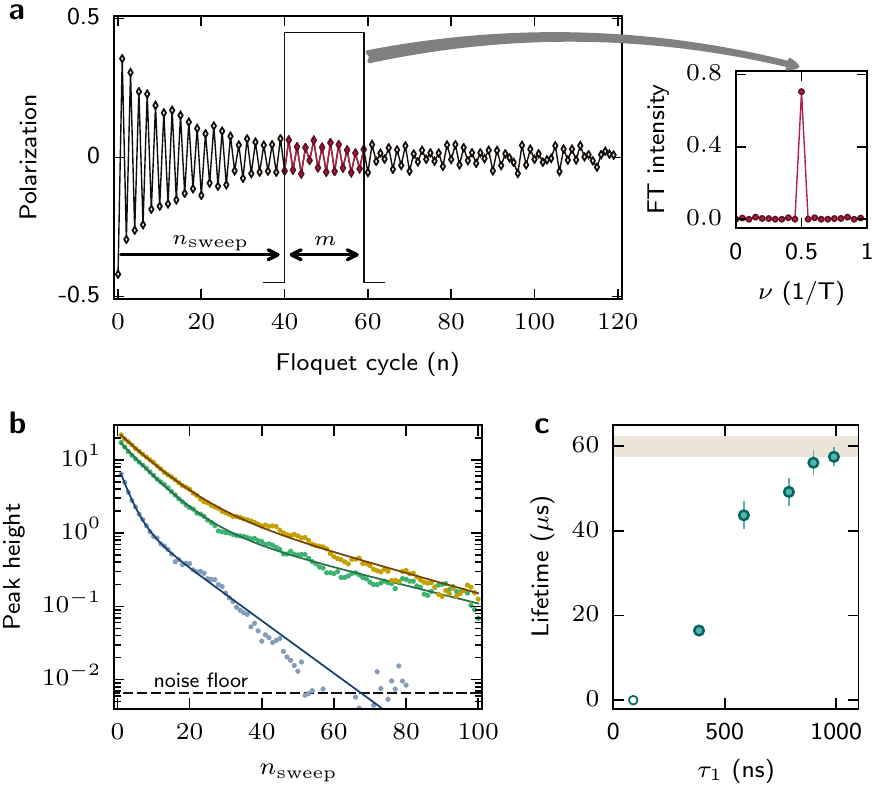}
\caption{\textbf{Long-time behavior of time-crystalline order.}
\textbf{a} Representative time trace of the spin polarization $P(nT)$ in the crystalline phase ($\tau_1=790 \,\mathrm{ns}$ and $\theta=1.034\pi$). The time-dependent intensity of the $\nu=1/2$ peak is extracted from a short-time Fourier transformation with a time window of length $m=20$ shifted from the origin by $n_{\mathrm{sweep}}$. \textbf{b} Peak height at $\nu=1/2$ as a function of $n_{\mathrm{sweep}}$ for three different pulse imperfections, $\theta=1.00\pi$ (yellow), $\theta=1.034\pi$ (green) and $\theta=1.086\pi$ (blue) ($\tau_1=790 \,\mathrm{ns}$).  
Lines indicate fits to the data using a phenomenological double exponential function. The noise floor corresponds to 0.017, extracted from the mean value plus the standard deviation of $\sum_\nu |S(\nu)|^2$ excluding the $\nu$ = 1/2 peak.
\textbf{c} Extracted lifetime of the time-crystalline order as a function of the interaction time $\tau_1$, for $\theta = 1.034\pi$.
Shaded region indicates the spin life-time $T_1^{\rho} = 60 \pm2~\mu$s due to coupling with the external environment.  The vertical error bars display the statistical error (s.~d.) from the fit. 
}
\label{fig:Figure2}
\end{figure}

The robustness of DTC order at late times is further explored in Fig.~2. 
With an interaction time $\tau_1 = 790~$ns and $\theta = 1.034\pi$,
the polarization exhibits an initial decay followed by persistent oscillations over the entire time window of our experimental observations (Fig. 2a). 
We perform a Fourier transform on sub-sections of the time-trace with a sweeping window of size $m=20$ (Fig. 2a) and extract the intensity of the $\nu=1/2$ peak as a function of the sweep position, $n_\textrm{sweep}$ (Fig. 2b). 
The $\nu=1/2$ peak intensity clearly exhibits two distinct decay timescales.
At short times, we observe a rapid initial decay corresponding to non-universal dephasing dynamics, while at late times, we observe a slow decay indicative of the persistence of DTC order.
Interestingly, the long-time decay rate seems relatively insensitive to the change of $\theta$ from $\pi$ to $1.034\pi$, and to variations in the initial spin states (see Methods), but significantly increases as one approaches the DTC phase boundary near $\theta = 1.086\pi$.
We fit the slow decay to an exponential to extract a lifetime for the time-crystalline order. As shown in Fig.~2c, for $\theta = 1.034\pi$,
this lifetime increases with the effective interaction strength (captured by $\tau_1$) and eventually approaches the independently measured spin depolarization time $T_1^\rho \sim~60~\mu\mathrm{s}$. This
demonstrates that for sufficiently strong interactions, the observed DTC order is only limited by coupling to the environment~\cite{choi2016depolarization}.

\begin{figure}[ht]
\centerline{\includegraphics[width=76mm]{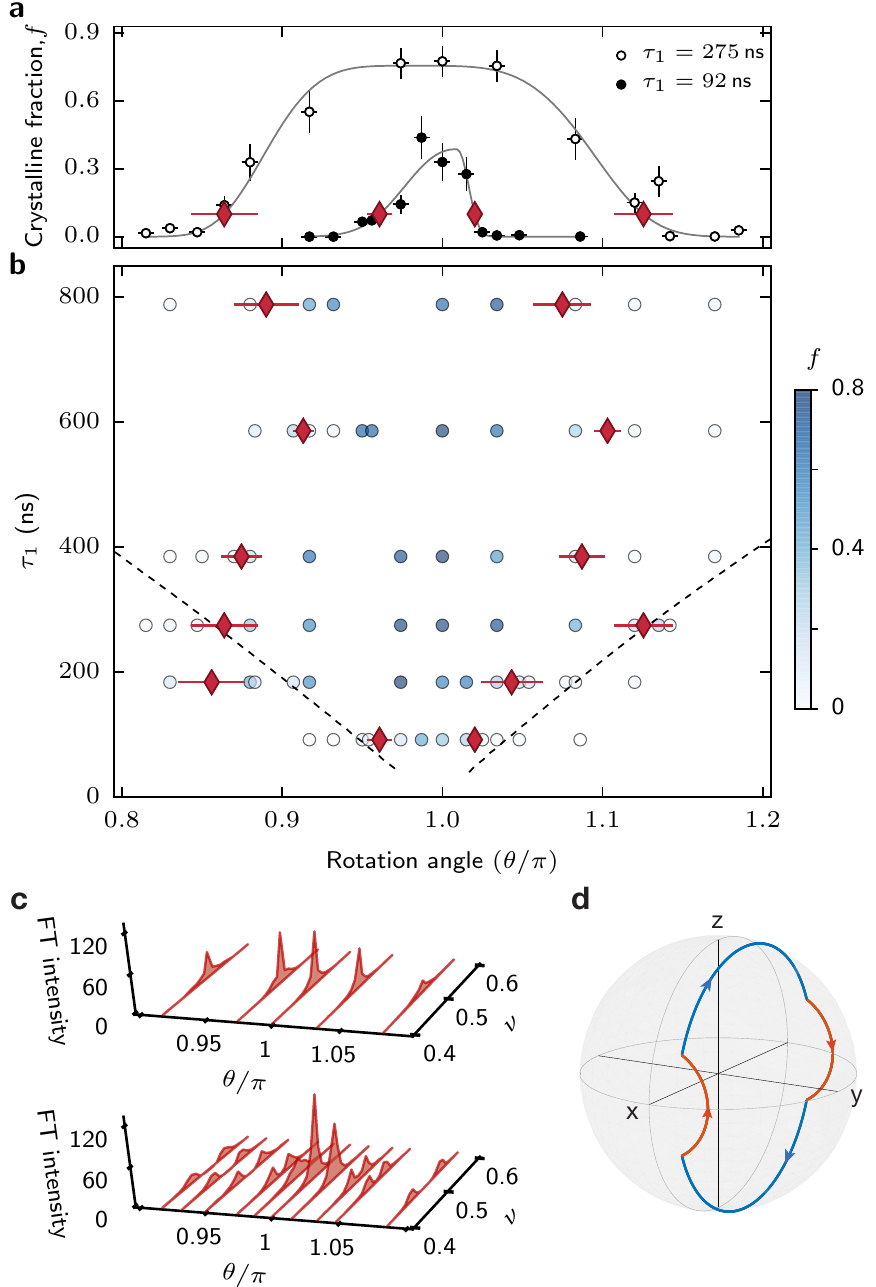}}
\caption{\textbf{Phase diagram and transition.}
\textbf{a} Crystalline fraction $f$ as a function of $\theta$ obtained from a Fourier transform at late times ($50< n \leq 100$).
Vertical error bars are limited by the noise floor (see Methods), horizontal error bars indicate the pulse uncertainty of $1\%$. 
Grey lines denote a super-Gaussian fit to extract the phase boundary (see Methods).
In  \textbf{a, b}, red diamonds mark the phenomenological phase boundary, identified as a $10\%$ crystalline fraction. Horizontal error bars denote the statistical error (s.~d.) from the fit. 
 The colors of the round data points in  \textbf{b} represent the extracted crystalline fraction at the associated parameter set.
 The dashed line corresponds to a disorder-averaged theoretical prediction for the phase boundary. Asymmetry in the boundary arises from an asymmetric distribution of rotation angles (see Methods).
 \textbf{c} Evolution of the Fourier spectra as a function of $\theta$ for two different interaction times, $\tau_1=385\,$ns (top) and $\tau_1=92\,$ns (bottom). 
 \textbf{d} Bloch sphere indicating a single spin trajectory of the $2T$-periodic evolution under the long-range dipolar  Hamiltonian (red) and global rotation (blue). 
}
\label{fig:Figure3}
\end{figure}

To experimentally determine the DTC phase boundary, we focus on the long-time behavior of the polarization time traces ($50<n\leq 100$) and compute the ``crystalline fraction'' defined as the ratio of the $\nu = 1/2$ peak intensity to the total spectral power, $f=|S(\nu=\frac{1}{2})|^2/\sum_{\nu} |S(\nu)|^2$ (see Methods). Figure 3a shows $f$  as a function of $\theta$ for two different interaction times.
For weak interactions ($\tau_1 = 92$~ns),  $f$ has a maximum at $\theta = \pi$ but rapidly decreases as $\theta$ deviates by $\sim 0.02 \pi$.
However, for stronger interactions ($\tau_1 = 275$~ns), we observe a robust DTC phase which manifests as a large crystalline fraction over a wide range $0.86\pi<\theta<1.13 \pi$.
We associate a phenomenological phase boundary with $f = 10\%$ and observe that the boundary enlarges with $\tau_1$, eventually saturating at $\tau_1 \approx 400\,$ns (Fig. 3b).
The phase boundary can also be visualized as the vanishing of the $\nu=1/2$ peak and the simultaneous emergence of two incommensurate peaks (Fig.~3c).
%

The rigidity of the $\nu=1/2$ peak can be qualitatively understood by constructing effective eigenstates of $2T$ Floquet cycles. 
We approximate the unitary time evolution over a single period as $U_T = R_y^\theta e^{-iH_{\rm eff} \tau_1}$ and solve for a self-consistent evolution using product states as a variational ansatz.
To this end, we consider the situation where a typical spin returns to its initial state after $2T$: $ \ket{\psi(0)} \propto \ket{\psi (2T)} = e^{-i\theta S^y} e^{i \phi_i S^x }
    e^{-i\theta S^y} e^{-i \phi_i S^x } \ket{\psi(0)}$, and self-consistently determine the interaction-induced rotation angle $\phi_i \equiv \sum_j J_{ij}/r_{ij}^3 \langle S_j^x \rangle \tau_1 \approx {\bar J}_i \tau_1 \bra{\psi(0)} S^x \ket{\psi(0)}$, 
where $\ket{\psi(0)}$ is the initial spin state and ${\bar J}_i = \sum_j J_{ij}/r_{ij}^3$ (see Methods). 
One expects $\phi_i$ to change sign after each Floquet cycle, since the average polarization $\bra{\psi(0)} S^x \ket{\psi(0)}$ should be flipped.
Intuitively, the self-consistent solution can be visualized as a closed path on the Bloch sphere (Fig.~3d), where each of the four arcs corresponds to one portion of the $2T$ periodic evolution.
When $\theta = \pi$, such a solution always exists.
More surprisingly, even when $\theta \neq \pi$, a closed path can still be found for sufficiently strong interactions, $|\bar{J}_i \tau_1| > 2|\theta_i - \pi |$; in such cases, the deviation in $\theta$ away from $\pi$ is compensated  by the dipolar interactions (Fig.~3d).
We obtain a theoretical phase boundary by numerically averaging the self-consistent solution over both disordered spin positions and local fields. The resultant phase boundary is in reasonable agreement with the experimental observations for short to moderate interaction times $\tau_1$, but overestimates the boundary at large $\tau_1$ (dashed line, Fig. 3b, see Methods).

\begin{figure}[ht]
\includegraphics[width=76mm]{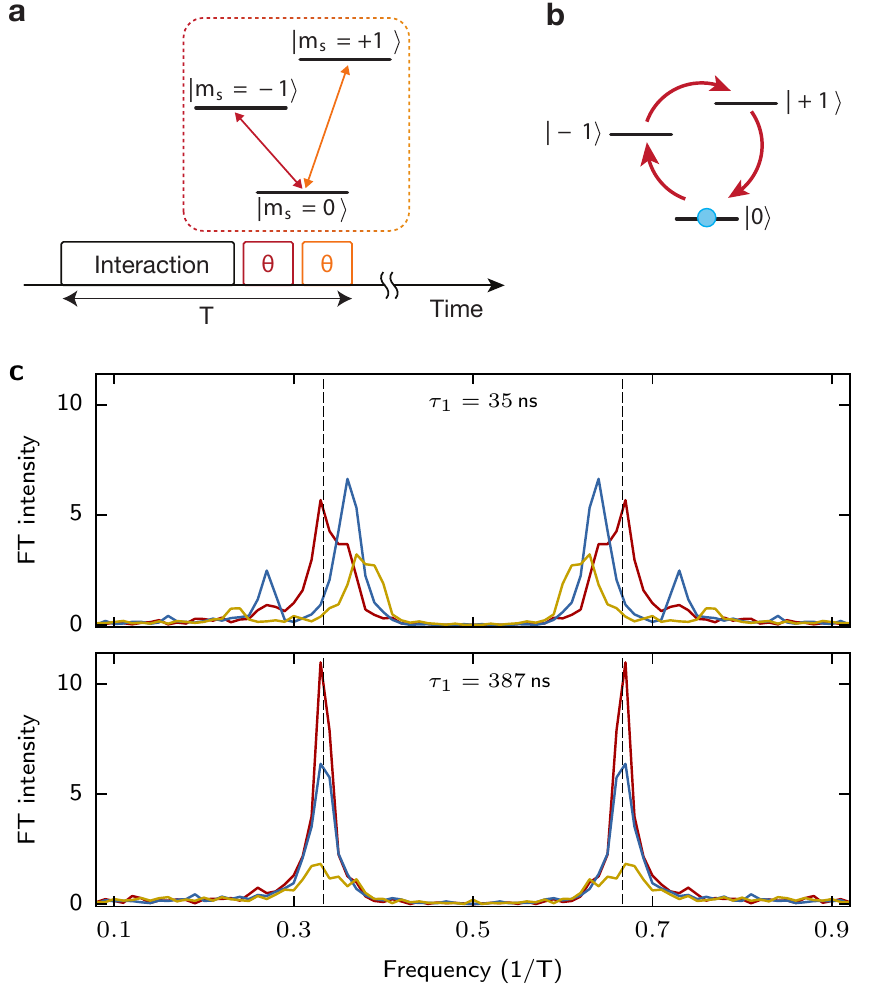}
\caption{\textbf{$\mathds{Z}_3$ time-crystalline order.} 
\textbf{a} Experimental sequence to demonstrate a $3T$-periodic discrete time-crystalline order. A single Floquet cycle is composed of three operations: time evolution under long-range dipolar Hamiltonian and rapid microwave pulses for two different transitions. 
\textbf{b} Visualization of the $3T$-periodicity in the polarization dynamics for the case of $\theta = \pi$. 
\textbf{c} Fourier spectra of the polarization dynamics for two different interaction times and for three different rotation angles $\theta$: $1.00\pi$ (red), $1.086\pi$ (blue) and $1.17\pi$ (yellow). 
Dashed lines indicate $\nu =\pm 1/3$.
}
\label{fig:Figure4}
\end{figure}

%
Finally, Fig.~4  demonstrates that the discrete time-translation symmetry can be further broken down to $\mathds{Z}_3$~\cite{von2016phase2,yao2016discrete,PhysRevB.94.045127,else2016floquet,von2016absolute}, resulting in DTC order at $\nu=1/3$.
Here, we utilize all three spin states of the NV center. 
We begin with all spins polarized in the $\ket{m_s=0}$ state and evolve under the bare dipolar Hamiltonian for a duration $\tau_1$ (see Methods).  
Next, we apply two resonant microwave pulses, each of duration $\tau_2$, first on the transition $\ket{m_s=0} \rightarrow\ket{m_s=-1}$ and then on the transition $\ket{m_s=0} \rightarrow \ket{m_s=+1}$.
In combination, this sequence of operations defines a single Floquet cycle with period $T = \tau_1 + 2 \tau_2$. As before, we measure the polarization, $P(nT)$, defined as the population difference between the $\ket{m_s=0}$ and $\ket{m_s=-1}$ states  (Fig.~4a).
When each of the applied microwaves corresponds to an ideal $\pi$-pulse, this sequence realizes a cyclic transition with $\mathds{Z}_3$ symmetry (Fig.~4b), which is explicitly broken by any change in the pulse duration.
 The Fourier spectra of $P(nT)$ for various pulse durations and two different values of $\tau_1$ are shown in Fig.~4c.
With weak interactions ($\tau_1 = 35$~ns), the position of the peaks is extremely sensitive to perturbations, but with sufficiently strong interactions ($\tau_1 = 387$~ns) the peaks are pinned to a rigid value of $\nu = 1/3$ despite perturbations as large as $17\%$, indicating the observation of $\nu=1/3$ DTC order.
Our observation of DTC order cannot be simply explained within current theoretical frameworks based upon either localization \cite{vedika2016phase,else2016floquet,von2016absolute,yao2016discrete} or  pre-thermalization~\cite{abanin2015exponentially,else2016pre}. 
In particular,  the present system with long-range dipolar interactions is not expected to be localized in either the static or the driven cases.
 In the static case, it has been previously demonstrated that it exhibits slow thermalization associated with critical dynamics \cite{kucsko2016critical}. 
In the driven case, the  long-time evolution is governed by the average Hamiltonian $D \simeq \sum_i (J_{ij}/r_{ij}^3) S_i^x S_{j}^x  + (\theta - \pi)/T \sum_i  S_i^y$, which likewise does not yield localized dynamics~\cite{anderson1958absence, yao2014many}.
We further note that the effective Hamiltonian of the $\mathds{Z}_3$ DTC includes not only Ising-type interactions but also spin exchange interactions, providing additional channels for thermalization (see Methods).

In principle, even in the absence of localization, time-crystalline order can persist for a long, but finite, pre-thermal time-scale \cite{else2016pre,abanin2015exponentially}.
Within this time-scale,  the spin system relaxes to a pre-thermalized state, defined as the thermal ensemble of $D$ with a  temperature determined by the energy density of the initial state.
Since our initially polarized state is effectively at infinite temperature with respect to $D$ (owing to the random signs of the dipolar couplings), one does not expect to observe pre-thermal DTC order. 
This is in stark contrast to our actual observations, which show that the DTC lifetime is  limited \emph{only} by the depolarization time $T_1^\rho$ due to coupling with the environment (Fig. 2c). We have explicitly verified that the DTC order is not significantly affected by varying the initial polarization (see Methods). 
One possible explanation is that due to slow critical thermalization \cite{kucsko2016critical}, the spins in our system do not reach even a pre-thermal state.
Such a critical regime of DTC  requires further theoretical investigation. 
Finally, the possibility that periodic driving itself can induce localization cannot be ruled out.

A number of remarkable phenomena in quantum dynamics have recently been observed in engineered many-body systems consisting of  ten to a few hundred particles~\cite{langen2015experimental,schreiber2015observation,monroe2016MBL,kaufman2016quantum,Bohnet1297}.
Our present observations indicate that robust DTC order can occur in large systems without fine-tuned interactions and disorder, even in the regime where localization is nominally not expected to occur. Beyond raising important questions about the role of localization and long-range interactions in studies of driven systems, our work opens up several new avenues for fundamental studies and potential applications. In particular, it should be possible to extend these studies 
to explore novel, dynamical quantum phases in more complex driven Hamiltonians. 
It is interesting to explore if such novel phases can be used to create and stabilize coherent quantum superposition states for  applications such as quantum metrology \cite{deutsch2010spin,anaMaria2008,cappellaro2009quantum}.

{\emph Note}: During the preparation of this manuscript, we became aware of a closely related work \cite{monroe2016time_crystal} where DTC order was observed in a system of 10 trapped ions.

\textbf{Acknowledgements} 
We thank A.Vishwanath, S. L. Sondhi and M. Zaletel
for insightful discussions and N. P. De Leon and P. C. Maurer for fabricating the diamond nanobeam and experimental help.
This work was supported in part by CUA, NSSEFF, ARO MURI, Moore Foundation, Harvard Society of Fellows, Princeton Center for Theoretical Science, Miller Institute for Basic Research in Science, Kwanjeong Educational Foundation, Samsung Fellowship, NSF PHY-1506284, NSF DMR-1308435, Japan Society for the Promotion of Science KAKENHI (No. 26246001), EU (FP7, Horizons 2020, ERC), DFG, SNF, Volkswagenstiftung and BMBF.

\textbf{Author contributions} 
S.C. and M.D.L. developed the idea for the study. J.C., R.L. and G.K. designed and conducted  the experiment. H.S., S.O., J.I. and F.J. fabricated the sample. S.C., H.Z., V.K., C.V., N.Y. and E.D. conducted theoretical analysis. All authors discussed the results and contributed to the manuscript.

\bibliography{TimeCrystal_ref}

\clearpage
\newpage

\section*{Methods}

\subsection{Experimental details}
Our sample and experimental setup have been previously described~\cite{kucsko2016critical}.
We utilize a diamond sample containing a high concentration ($\sim 45$~ppm) of NV centers, corresponding to an average NV-NV separation of $ 5~$nm. For a single crystalline orientation of NV centers, selected by applying an external magnetic field, this corresponds to an average separation of $ 8~$nm, resulting in a typical dipolar interaction strength of $2\pi\times 105$~kHz.
The system furthermore exhibits strong on-site energy disorder, owing to the effects of lattice strain, the random position of NV centers as well as the presence of scattered paramagnetic impurities (consisting mainly of P1 centers and $^{13}$C nuclear spins).
For each NV, the effective random field $\Delta_i$ is therefore a function of its local environment, including interaction effects of neighboring NV centers. This results in an approximately Gaussian distribution with standard deviation $W = 2\pi\times 4.0$~MHz. We extract $W$ by measuring the linewidth of an ESR spectrum with sufficiently weak microwave driving strength to avoid power broadening.
In order to control the experimental probe volume, we fabricate a diamond nanobeam structure ($\sim300$~nm$~\times~300$~nm$~\times~20~\mu$m) and confocally address a region of $\sim 300$~nm diameter using a green laser ($532$~nm). This realizes an effective three dimensional excitation volume containing $\sim10^6$ NV centers.
By applying an external magnetic field along one of the diamond crystal axes, we spectrally isolate one group of NV centers and selectively address an effective two-level system between the $\ket{m_s = -1}$ and $\ket{m_s=0}$ spin states via coherent microwave radiation. The addition of a microwave IQ-mixer allows for arbitrary rotations around any linear combination $\hat{x}$ and $\hat{y}$.
 
\subsection{Experimental sequence}
Initial polarization of NV centers into $\ket{m_s=0}$ is performed via laser illumination at a wavelength of 532~nm, a power of $50~\mu \textrm{W}$ and a duration of 20~$\mu$s. Subsequent application of a microwave ($-\pi/2$)-pulse along the $\hat{y}$ axis is used to coherently rotate the spin ensemble into $\ket{+} = (\ket{m_s=0}+\ket{m_s=-1})/\sqrt{2}$.
The spins are then subjected to continuous driving at a Rabi frequency $2\pi\times 54.6\,$MHz along the $\hat{x}$ axis for a duration $\tau_1$.
This so-called spin-locking technique suppresses two-spin (flip-flip and flop-flop) processes due to energy conservation as well as to decouple spins from their environment~\cite{kucsko2016critical}. In our sample, this technique leads to spin lifetimes of $\sim 60~\mu$s~\cite{choi2016depolarization}. Finally, we apply a short microwave pulse along the $\hat{y}$ axis over an angle $\theta \sim \pi$.
We repeat this Floquet cycle with various values of $\theta$, controlled by changing the Rabi driving strength as well as the pulse duration.
The imperfection in microwave manipulations (for initialization into $\ket{+}$ as well as rotation angles $\theta$) amounts to $\sim1\%$, arising from a combination of spatial inhomogeneity of the driving field (0.5\%) as well as on-site potential disorder (0.9\%). Following a coherent time evolution, the spin state of the NV ensemble is optically detected by applying a final ($\pi/2$)-pulse along the $\hat{y}$ axis and measuring the population difference in the $\ket{m_s=0}$ and $\ket{m_s=-1}$ basis. The polarization is defined as $P=\ket{m_s=0} - \ket{m_s=-1}$ by calibrating the NV fluorescence using a Rabi oscillation contrast measurement. To avoid heating of the sample, resulting in  drifts in the Rabi frequency, a waiting time of $600-900\,\mu$s is implemented before the sequence is repeated. The minimum spacing between microwave pulses is maintained at $1~$ns.

To understand the effect of different initial states on the DTC phase, we replaced the initial $(-\pi/2)$-pulse with a $(-\pi/3)$-pulse. This results in the preparation of a global spin state, which is rotated from the $\hat{x}$ axis by $\pi/6$. Despite this change, the measured DTC lifetime ($47.6\pm2.4\, \mu$s) agrees well with that of the polarized spin state  ($49.2\pm3.3\,\mu$s), demonstrating that  DTC order is insensitive to the initial state. 

\subsection{Experimental identification of phase boundary}

To identify the position of the phase boundary in our experiment, we define the crystalline fraction $f$ as $f = |S(\nu=\frac{1}{2})|^2/\sum_{\nu} |S(\nu)|^2$. Error bars in $f$ are calculated via error propagation in consideration of the noise floor in the Fourier spectrum; each measured spectrum contains a background noise level $\sigma_n$, resulting in a variation of $f$ as,
\begin{widetext}
\begin{align}
\delta f = f \sqrt{\left(\sigma_n/ |S(\nu=\frac{1}{2})|^2\right)^2 + \left(N\sigma_n / \sum_{\nu} |S(\nu)|^2\right)^2 - 2N\sigma_n^2 /\left( |S(\nu=\frac{1}{2})|^2 \sum_{\nu} |S(\nu)|^2\right)}, 
\end{align}
\end{widetext}
where $N = 50$ is the number of points in the Fourier spectrum. This gives rise to an uncertainty in the DTC fraction: $f \in [f-\delta f, f+\delta f]$ (Fig. 3a). 
To extract the phase boundary, we use a phenomenological, super-Gaussian function
\begin{align}
   F_{\tau_1} (\theta)
   = \left\{ 
    \begin{array}{cc}
    f_{\tau_1}^{\text{max}}\exp  \left[ -\frac{1}{2} \left( \frac{\abs{\theta - \theta_0}}{\sigma_-} \right)^p  \right] & \text{, } \theta \leq \theta_0 \\
    f_{\tau_1}^{\text{max}}\exp  \left[ -\frac{1}{2} \left( \frac{\abs{\theta - \theta_0}}{\sigma_+} \right)^p  \right] & \text{, } \theta \geq \theta_0 \\
    \end{array}
    \right.
\end{align}
where $\sigma_{\pm}$, $\theta_0$, $p $ are the characteristic width, central position and the power of the super-Gaussian fit, and $f_{\tau_1}^\text{max}$ is the maximum value of the DTC fraction for a given duration $\tau_1$. The proposed function naturally captures the observed asymmetry in the phase boundary. We define the phase boundary as the rotation angle $\theta_{\pm}$ where $F_{\tau_1}(\theta_{\pm})$ = 0.1, i.e. $\theta_\pm= \theta_0 \pm \sigma_\pm\left[2\ln(f_{\tau_1}^{\max}/0.1)\right]^\frac{1}{p}$. Errors in the phase boundary are derived from the fit uncertainties. 

\subsection{Theoretical description}
As a variational ansatz, we consider the time evolution of a homogeneous product state of the form $\ket{\Psi} = \ket{\psi_0}^{\otimes N}$ with $\ket{\psi_0} = \cos(\theta_0/2) \ket{+} + \sin(\theta_0 /2 ) e^{i\phi_0} \ket{-}$, where $\ket{\pm} = (\ket{m_s=0}\pm\ket{m_s=-1})/\sqrt{2}$.
The qualitative behavior does not change even if we allow spins to be oriented in different directions.
An approximate eigenstate for the time evolution over two periods is obtained by solving the equation for a single spin, $\ket{\psi_0} = e^{-i\theta S^y} e^{i \phi_i S^x } e^{-i\theta S^y} e^{-i \phi_i S^x } \ket{\psi_0}$ with a self-consistently determined $\phi_i = \bar{J}_i \bra{\psi_0}S^x\ket{\psi_0}$ where $\bar{J}_i = \sum_{j} J_{ij}/r_{ij}^3 $ is the total strength at site $i$.
The sign of $\phi_i$ is flipped in the second evolution as the spin polarization along the $\hat{x}$ direction alternates in each cycle.
Note that we have ignored the effects of the on-site disorder potential $\Delta_i$, interactions during global rotations and rotations induced by $\Omega_x$. This is justified due to the high microwave driving strength $\Omega_{x(y)} \gg W$ and $\Omega_x \tau_1$ being integer multiples of $2\pi$. (The effects of on-site disorder are fully included in the numerical computations.)
A non-trivial solution ($\theta_0 \neq \pm \pi$) is obtained if the first two rotations result in a vector that is rotated by $\pi$ along the $\hat{y}$~axis (Fig.~3d), which is satisfied when
$\phi_0 = m\pi -\phi_i/2$ with $m \in \mathds{Z}$ and $\cot\theta_0 = - (-1)^m \tan(\theta/2) \sin(\phi_i/2)$.
Solving for $\cos^2 \theta_0$ yields 
\begin{align}
    \cos^2 \theta_0 &= \frac{ \tan^2(\theta/2) \sin^2 (\phi_i/2)}
    {1+\tan^2(\theta/2) \sin^2 (\phi_i/2) }.
\end{align}
Using $\phi_i = \bar{J}_i \tau_1 \cos \theta_0$, one can show that a solution exists only when $|\tan{(\theta/2)} \bar{J}_i \tau_1 /4| > 1$, implying that $|\theta-\pi| < | \bar{J}_i \tau_1/2|$ in the vicinity of $\theta \approx \pi$.

The linear dependence of the phase boundary is consistent with the phase diagram provided in Ref.~\cite{vedika2016phase,yao2016discrete}.
As long as a solution exists, small variations in $\theta$ correspond to a smooth deformation of the closed trajectory. Therefore, the existence of such a closed path stabilizes the time-crystalline phase.
{\color{black} We emphasize that such a $2T$-periodic path is a consequence of interactions; without the change of sign in $\phi_i$, the eigenstates of the unitary evolution over one or two periods coincide, and therefore, unless the rotation angle is fine-tuned, $T$-periodic motion cannot be broken into a $2T$ period.}
The eigenstates of unitary evolution over one period can be obtained as  even and odd linear combinations, $(\ket{\Psi} \pm e^{-i\epsilon_i} U_1 \ket{\Psi})/\sqrt{2}$, where $U_1 = \otimes_i (e^{-i\theta S_i^y} e^{-i \phi_i S_i^x })$, and the quasi-energy eigenvalue is given by $e^{i2\epsilon_i} = \bra{\Psi} (U_1)^2 \ket{\Psi}$. %

To estimate the phase boundary,
we numerically solve the self-consistency equation.
Here, we include the effects of on-site disorder potential $\Delta_i$ in all four rotations as well as the disorder in $\bar{J}_i$ arising from the random positions of NV centers.
The distribution of $\bar{J}_i$ is simulated for 1000 spins, randomly distributed in three dimensions with an average separation $r_0$ and minimum cutoff distance $r_\textrm{min} = 3~$nm (limited by NV-NV electron tunneling~\cite{choi2016depolarization}).
Instead of $\cos(\theta_0)$, we solve for a self-consistent distribution for $\cos(\theta_0)$, where $\langle S^x\rangle$ is defined as the mean of the distribution.
The average order parameter $\langle \cos^2\theta_0\rangle$ is computed for various values of $\tau_1$ and $\theta$ and compared with a threshold value of 0.1 in order to identify the phase boundary.
The experimental and numerical phase boundaries are asymmetric about $\theta = \pi$.
We attribute this to the inherently asymmetric distribution of the effective rotation angle, $\theta_i \approx \tau_2\sqrt{\Omega_y^2 + (\Delta_i+ \bar{J_i} )^2}$, which causes the  transition to occur earlier for positive deviations $\theta-\pi$.

While we assumed $\phi_i$ to be a classical variable in this analysis, the interaction induced rotation angle is an operator $\hat{\phi}$ that exhibits quantum fluctuations and leads to non-trivial quantum dynamics.
Under such dynamics, spins get entangled, resulting in mixed state density matrices.
These effects cannot be ignored in the case of long interaction times,
effectively limiting the present description.
We believe that the diminished range of $\theta$ in the experimentally obtained phase diagram (Fig.~3b) is related to this effect.

\subsection{Derivation of Effective Hamiltonian for $\mathds{Z}_3$ symmetry breaking phase}
Using microwave driving resonant with two different transitions (Fig.~4a), we realize dynamics involving all three spin states and observe a robust $3T$-periodic time-crystalline order. The unitary matrix of the time evolution during the fundamental period $T$ is given as
\begin{align}
U_{3} = e^{-i \sum_i(\sigma^i_{-1,0}+\sigma^i_{0,-1}) \theta/2}e^{-i \sum_i(\sigma^i_{+1,0}+\sigma^i_{0,+1}) \theta/2}e^{-i H_2\tau},\nonumber
\end{align}
where $\sigma^i_{a,b} \equiv \ket{m_s=a}\bra{m_s=b}$ for spin-$i$ and $H_2 = H_\textrm{dis} + H_\textrm{int}$ is the effective Hamiltonian of NV centers for all three spin states including on-site disorder potentials $H_\textrm{dis} = \sum_i \Delta_i^+ \sigma^i_{+1,+1} + \Delta_i^- \sigma^i_{-1,-1}$ and dipolar interactions for spin-1 particles~\cite{kucsko2016critical}
\begin{align}
H_\textrm{int}=&
\sum_{ij} \frac{J_{ij}}{r_{ij}^3}
\left[
-
\frac{
\sigma^i_{+1,0}\sigma^j_{0,+1}+\sigma^i_{-1,0}\sigma^j_{0,-1}+h.c.}{2} \right. \nonumber \\
&+ \left. (\sigma^i_{+1,+1}-\sigma^i_{-1,-1})(\sigma^j_{+1,+1}-\sigma^j_{-1,-1})
\right].
\end{align}
We note that this Hamiltonian is obtained in the rotating frame under the secular approximation.
The Hamiltonian $H_2$ conserves the total population in any of the three spin states, $P_a = \sum_i \sigma^i_{aa}$ with $a \in \{ 0, \pm1\}$.
If each microwave pulse realizes a $\pi$-pulse ($\theta = \pi$), their combination results in a cyclic transition $R_3^\pi: \ket{m_s=+1} \mapsto -\ket{m_s=-1} \mapsto i\ket{m_s=0} \mapsto \ket{m_s=+1} $, and the population $P_0$ becomes periodic over three periods. Under such evolution, the effective Hamiltonian over three periods is given by $D_3^\pi = \left[ H_2 + (R_3^\pi)^{-1} H_2 R_3^\pi + (R_3^\pi)^{-2} H_2 (R_3^\pi)^2\right]/3$, in which on-site disorders average to zero, and the interactions are modified to
\begin{align}
D_3^\pi = \sum_{ij} \frac{J_{ij}}{r_{ij}^3}
\left[
\sum_{a} \sigma^i_{aa}\sigma^j_{aa} - \frac{1}{3} \sum_{a\neq b} \sigma^i_{ab}\sigma^j_{ba}
\right].
\end{align}
The first term describes Ising-like interactions that shift energy when any pair of spins are in the same state, and the second term corresponds to spin-exchange interactions that allow polarization transport.
For small perturbations in the microwave pulse angle $\epsilon = \theta-\pi$, the effective dynamics, to leading order, are governed by
\begin{align}
D_3^{\pi+\epsilon} \approx D_3^\pi + \frac{\epsilon}{3\tau} \sum_j \left(\sigma^j_{+1,0} + \sigma^j_{-1,0} + i \sigma^{j}_{+1,-1} + h.c.\right),\nonumber
\end{align}
which explicitly breaks the conservation laws for $P_a$. 

\end{document}